\documentclass[final,3p,times,twocolumn]{elsarticle}

\makeatletter
\def\ps@pprintTitle{%
  \let\@oddhead\@empty
  \let\@evenhead\@empty
  \def\@oddfoot{\reset@font\hfil\thepage\hfil}
  \let\@evenfoot\@oddfoot
}
\makeatother

\usepackage{graphicx}
\usepackage{dcolumn}
\usepackage{bm}
\usepackage{enumerate}

\usepackage{color}
\definecolor{DarkBlue}{rgb}{0.1,0.1,0.5}
\definecolor{Red}{rgb}{0.9,0.0,0.1}
\definecolor{boja1}{rgb}{0.5,0.3,0.7}
\definecolor{boja2}{rgb}{0.3,0.58,0.345}
\definecolor{boja3}{rgb}{0.631,0.62,0.216}

\usepackage{amssymb}
\biboptions{sort&compress}
\usepackage{float}

\def\be{\begin{equation}}
\def\ee{\end{equation}}
\def\bea{\begin{eqnarray}}
\def\eea{\end{eqnarray}}
\def\bfl{\begin{flushleft}}
\def\efl{\end{flushleft}}
\def\bfr{\begin{flushright}}
\def\efr{\end{flushright}}
\def\bc{\begin{center}}
\def\ec{\end{center}}
\def\ben{\begin{enumerate}}
\def\een{\end{enumerate}}
\def\bit{\begin{itemize}}
\def\eit{\end{itemize}}

\def\dzn{,\kern-0.1em,}

\def\d#1{{#1\kern-0.4em\char"16\kern-0.1em}}
\def\D#1{{\raise0.2ex\hbox{-}\kern-0.4em 31}}

\newcommand {\apgt} {\ {\raise-.5ex\hbox{$\buildrel>\over\sim$}}\ }
\newcommand {\aplt} {\ {\raise-.5ex\hbox{$\buildrel<\over\sim$}}\ }

\begin{document}

\begin{frontmatter}



\title{Thermodynamics of $O(3)$ Classical Heisenberg Model in Multipath Metropolis Simulation}


\author[pmf]{Petar Mali\corref{cor1}}
\ead{petar.mali@df.uns.ac.rs}

\author[pmf]{Slobodan Rado\v sevi\' c}
\ead{slobodan@df.uns.ac.rs}

\author[ftn]{Predrag S. Raki\' c}
\ead{pec@uns.ac.rs}

\author[ftn]{Lazar Stri\v cevi\' c}
\ead{lucky@uns.ac.rs}

\author[ftn] {Tara Petri\' c}
\ead{tara.petric@gmail.com}

\cortext[cor1]{Corresponding author}

\address[pmf]{Department of Physics, Faculty of Science, University of Novi Sad,
Trg Dositeja Obradovi\' ca 4, 21000 Novi Sad, Serbia}
\address[ftn]{Faculty of Technical Sciences, University of Novi Sad,
Trg Dositeja Obradovi\' ca 6, 21000 Novi Sad, Serbia}

\begin{abstract}
We study the thermodynamics of classical Heisenberg model using the multipath
approach to Metropolis algorithm Monte Carlo simulation.
This simulation approach produces uncorrelated results with known precision.
Also, it can be easily generalized to other classical models of magnetism.
Comparing results obtained from multipath and from single--path simulations
we demonstrate that these approaches produce equivalent results.
\end{abstract}


\end{frontmatter}

\section{Introduction}
Classical lattice models attract attention nowadays for several reasons. Classical Heisenberg model
is frequently used in Monte Carlo simulations of nonlinear sigma models \cite{Justin},
and also for modeling real compounds \cite{EuX,EuX2,Stenli} and other systems \cite{ptice,grain}.
In the recent paper \cite{nas} multipath  Metropolis simulation of 
$O(3)$ classical Heisenberg model is introduced. 
Since multipath approach is embarrassingly parallelizable, it utilizes
 easily 
 computing power of any number of computing elements and provides 
normally distributed results  with desired precision. 
 One of the main
 advantages of the multipath  Metropolis simulation is its  applicability 
to many different classical lattice models, 
such as Ising \cite{Montrol,San, Izing}, Potts \cite{Wu,Glumac} etc. 
The multipath approach allows complete control over the simulation in a sense that 
it is possible to conduct a ''short simulation''\footnote{Simulation with just a few 
simulation paths that can be conducted in short period of time.} 
in order to make a reasonable estimate.
Later, the simulation precision can be incrementally improved with additional,
subsequently computed results.
This is of great practical importance as it turns out that the optimal simulation parameters
(number of lattice sweeps and the number of simulation paths), 
strongly depend on the temperature and lattice size.

The simulation results presented in this paper were 
computed using free software C++ library called ''Hypermo'' \cite{web-Hypermo}
on computing services of the 
Supercomputing  Center of Galicia (CESGA) \cite{Cesga}. 
The figures are created using ''Tulipko'' \cite{web-tulip}
interactive visualization tool.

\section{Model and simulation}
The Hamiltonian of classical $O(3)$ Heisenberg model is
\begin{equation}
H=-\frac{J}{2}\sum_{\bm{n},\bm{\lambda}}\bm{S}_{\bm{n}}\cdot\bm{S}_{\bm{n}+\bm{\lambda}}
\label{ham},\end{equation}
 where the summation is taken over all lattice sites $\{\bm{n}\}$ with total $N=L^3$ sites of simple cubic lattice, and $\bm{\lambda}$ 
connects
a given site to its nearest neighbors. 
 The convinient
energy scale is set by $J=k_{\mathsf{B}}=1$ and 
we use the standard spherical parametrization for spin vectors
 \be \bm{S}_{\bm{n}}=[\sin \theta_{\bm{n}} \cos \varphi_{\bm{n}},\sin \theta_{\bm n}\sin \varphi_{\bm{n}},\cos \theta_{\bm{n}}]^{\mathsf{T}}.\ee

The quantities of  interest are the total spin
\be \bm{M}=\frac{1}{N}\sum_{\bm{n}}\bm{S_{\bm{n}}},\ee
of which the average value is magnetization $\langle \bm{M}\rangle$
,
the internal energy of the system $\langle H \rangle$, magnetic susceptibility
\be \chi(T)=\frac{L^3}{T} \left[\langle |\bm M|^2 \rangle-\langle |\bm M| \rangle^2 \right],\label{chi} \ee
 and capacity
 \be C_V(T)=\frac{L^3}{T^2}\left[\langle H^2 \rangle-\langle H \rangle^2 \right].\label{cv}\ee 
 Because there can't be no spontaneous symmetry breaking in finite
 lattices magnetic susceptibility is defined with
 \be |\bm{M}|=\frac{1}{N}\Big{|}\sum_{\bm{n}}\bm{S_{\bm{n}}}\Big{|}.\label{magnetic-susc}\ee
In multipath approach, each simulation consists of a certain number $\mathcal{N}$ of 
simulation paths (simulation path, SP). 
Each SP produces output.
Outputs of all the $\mathcal{N}$ SPs, together, form a simulation output (SO). Monte Carlo averages
are then computed as
\be \langle A \rangle=\frac{1}{\mathcal{N}}\sum_{i=1}^{\mathcal{N}}A_i \label{MC}\ee
and $\chi$ and $C_V$ are calculated from (\ref{chi}) and (\ref{cv}).
It should be noted that all thermodynamic quantities in the paper are
calculated per lattice site.

Multipath Metropolis simulation  can be easily visualized 
in the phase space of the lattice, which is
the direct product of the two-spheres $\mathcal{S}^{\mathbf 2}$ 
  located
at lattice sites\footnote{The state of each site is determined by 
two angles $\varphi \in[0,2\pi]$ 
and $\theta \in [0,\pi]$ and thus the
 dimension
of phase space is $\dim(PS)=2^{L^3}$.}.
\begin{figure}[ht] 
\includegraphics[width=7cm]{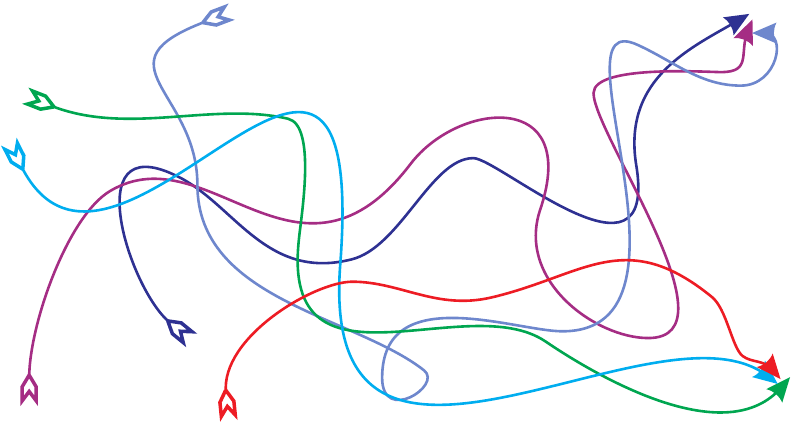}
\caption{(Color online) Illustration of the lattice phase space trajectories  
in the multipath simulation at low temperature for random initial state.
Each line represents one path.}
\label{Fig:multipath}
\end{figure}
Figure \ref{Fig:multipath} illustrates
multipath simulation in the lattice phase space ($PS$) at
low temperatures and random initial state. 
Every curve represents a single--path through the lattice phase space. 
Each path starts from some random state of the lattice and it contributes
with single result (the final state of that path) in (\ref{MC}).
In contrast to single--path simulation, there is no correlation 
between the multipath SP outputs.
Thus, standard statistical analysis can be applied on it (See \cite{nas}
for detail discussion). 
Note that existence of two limit points
in phase space is a consequence of finite lattice size \cite{nas,Binder81}.
\begin{figure}[H] 

\includegraphics [width=7.5cm]{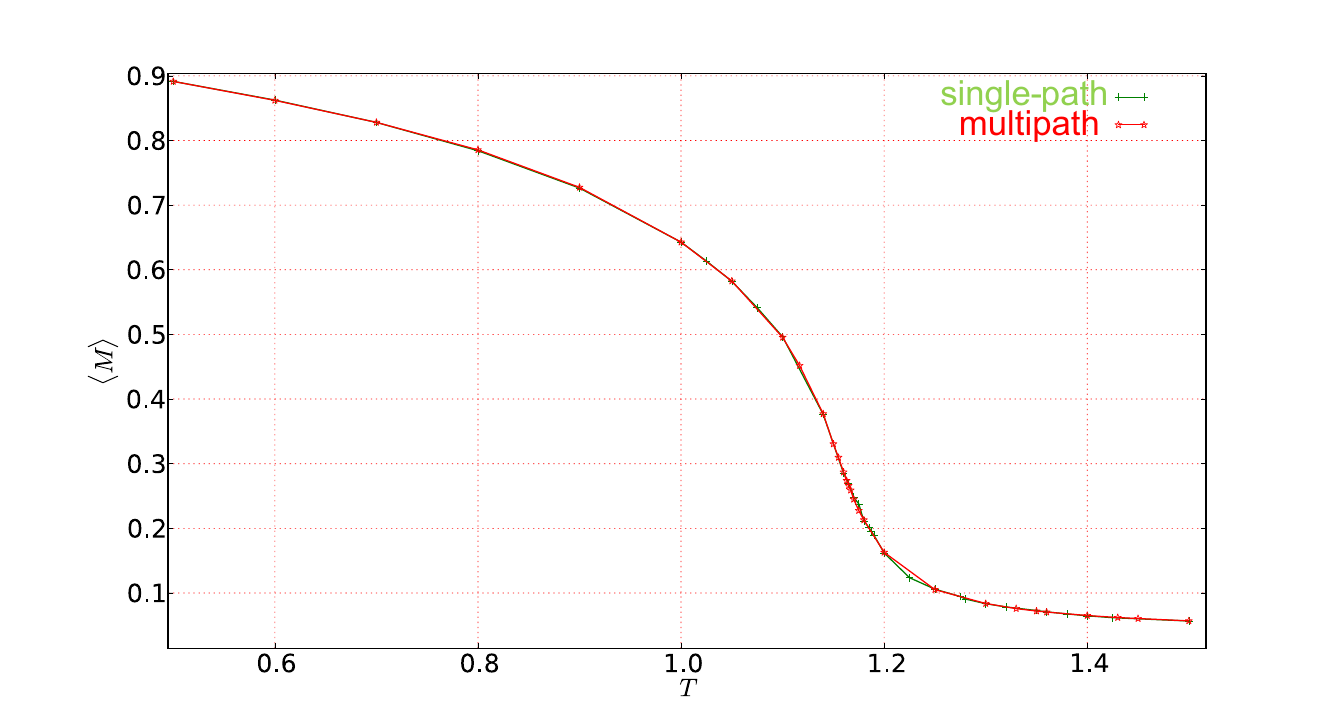}
\centering
\caption{(Color online) Magnetization as a function of temperature for $L=10$ in the single--path
and multipath approach. }
\label{Fig:mag}
\end{figure}

\section{Results and discussion}
 All simulations were conducted for linear
 size of the system
$L=10$ with
periodic boundary condition, in both single and multipath approach.
In single--path approach we used $2\times 10^6$ lattice sweeps to achieve
thermal equilibrium in whole temperature range, and afterwards only one out
of every five lattice sweeps was used to calculate the averages of
physical quantities \cite{Kitaev}. At every temperature $5 \times 10^5$ measurements were averaged.

To make sure that revailable results are generated by multipath simulation, it is prepared
in two different setups. In the first one, refered to as random initial state simulation in the text,
at every lattice site both angles $\theta$ and $\varphi$ are taken to be arbitrary. In the second one,
denoted as ordered initial state simulation spins are taken to points along z-axis, with no restriction
on second spherical angle $\varphi$.

We have to bear in mind, however, that multipath simulations naturally split
 into three temperature domains in which different numbers of lattice sweeps/simulation paths are needed.
In low temperature region for simulation convergence
(See \cite{nas}) more lattice sweeps is needed since all paths start
from some random state of the lattice.
(Simulation speed can be optimized if 
ordered state is taken to be ''starting point'' of all paths.) 
On the other hand, high temperature region
 requires more simulation paths. In the critical region
we take sufficiently large number of lattice sweeps and results due
to overlaping of the two different output distributions \cite{Binder81}.
\begin{figure}[H] 
\centering
\includegraphics [width=7.5cm]{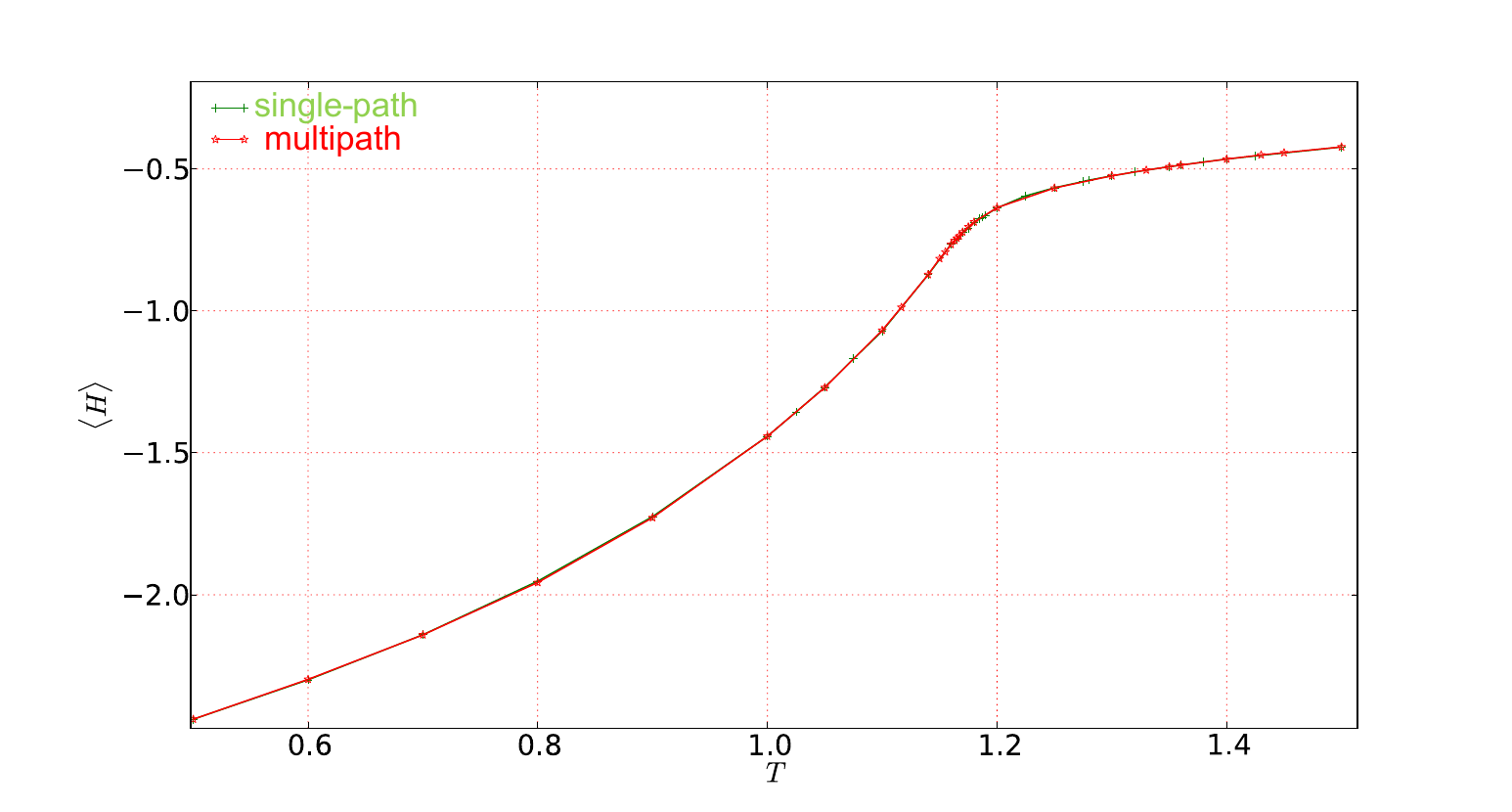}
\caption{(Color online) Energy as a function of temperature for 
$L=10$ in the single--path and multipath approach. }
\label{Fig:e}
\end{figure}
\begin{figure}[H] 
\centering
\includegraphics [width=6.0cm]{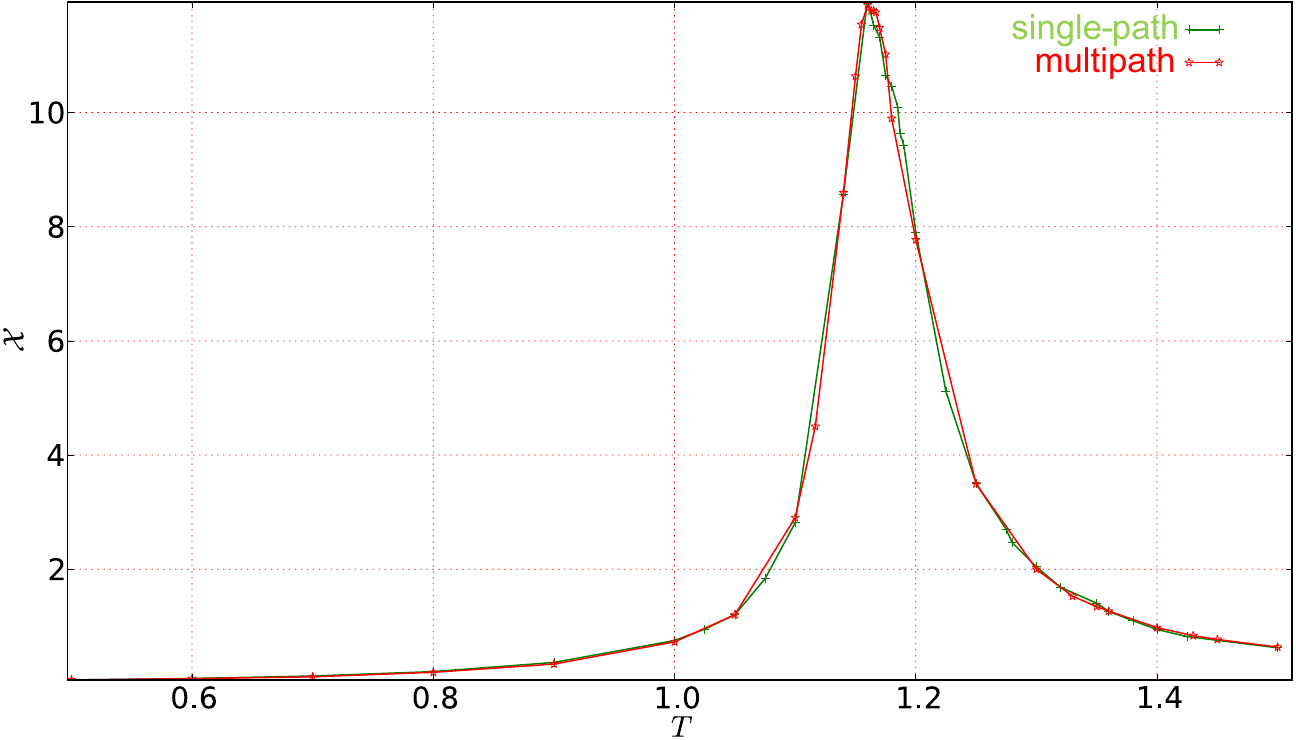}
\caption{(Color online) Magnetic susceptibility as a function of temperature for 
$L=10$ in the single--path and multipath approach. }
\label{Fig:susc}
\end{figure}
\begin{figure}[H] 
\centering
\includegraphics [width=6.0cm]{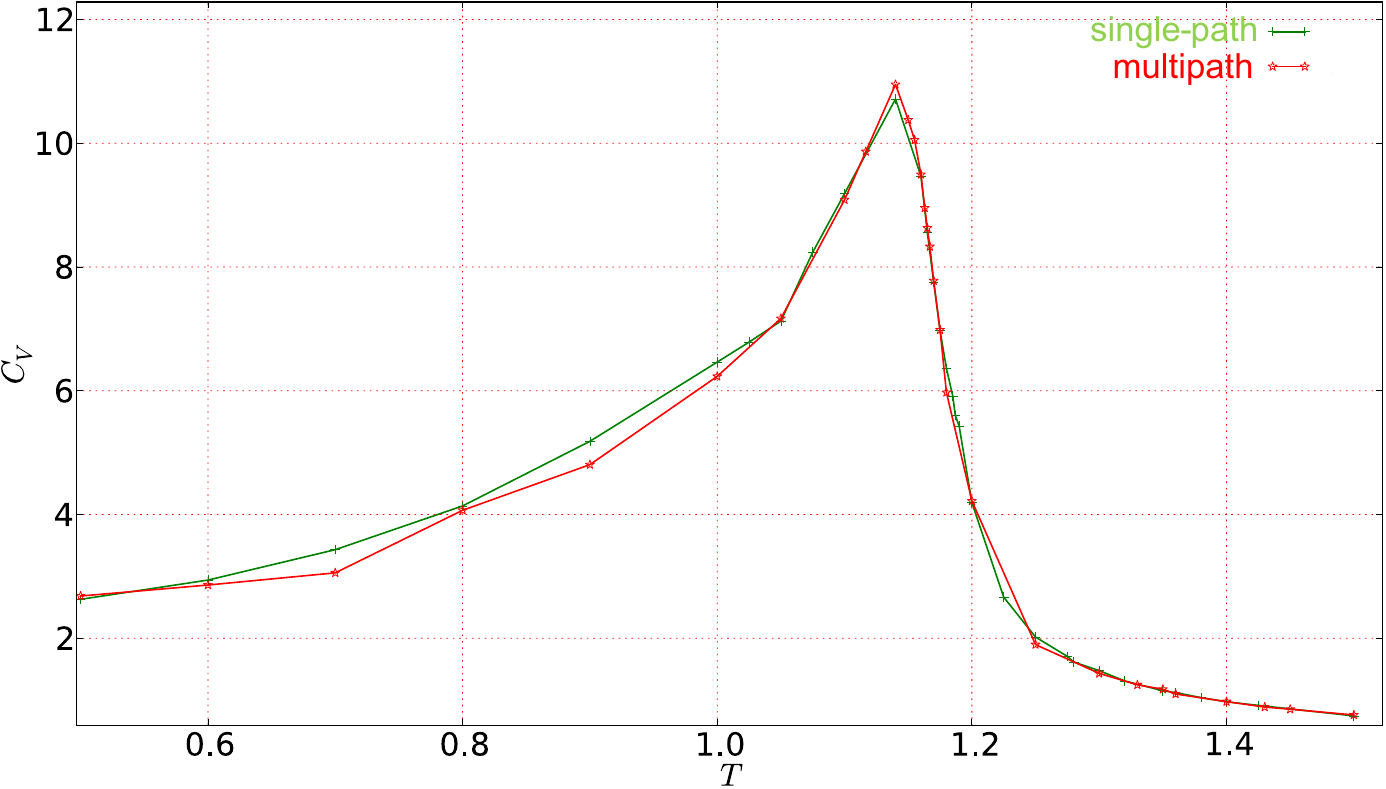}
\caption{(Color online) Heat capacity as a function of temperature for 
$L=10$ in the single--path and multipath approach. }
\label{Fig:tcap}
\end{figure}

From Figs. 2--5, we note that the differences in the thermodynamical characteristic
obtained by single--path and multipath approach are negligible. 
\begin{figure}[H] 
\includegraphics [width=5.0cm]{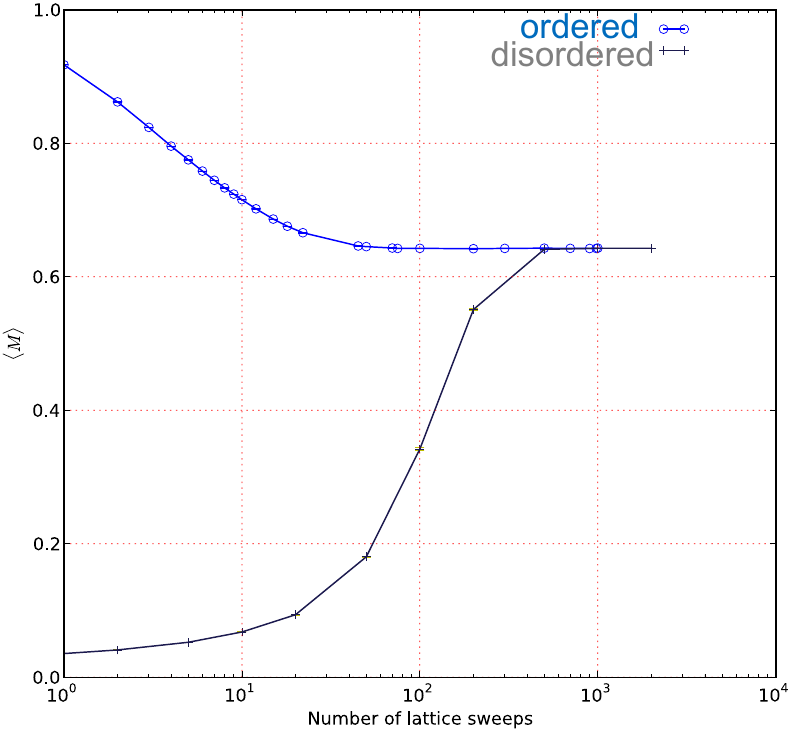}
\centering
\caption{(Color online) Magnetization, calculated starting
from both ordered and disordered states,  as a function
of number of lattice sweeps for $\mathcal{N}=10^4$, at $T=1$.}
\label{Fig:ee}
\end{figure}
\begin{figure}[H] 
\includegraphics [width=8.5cm]{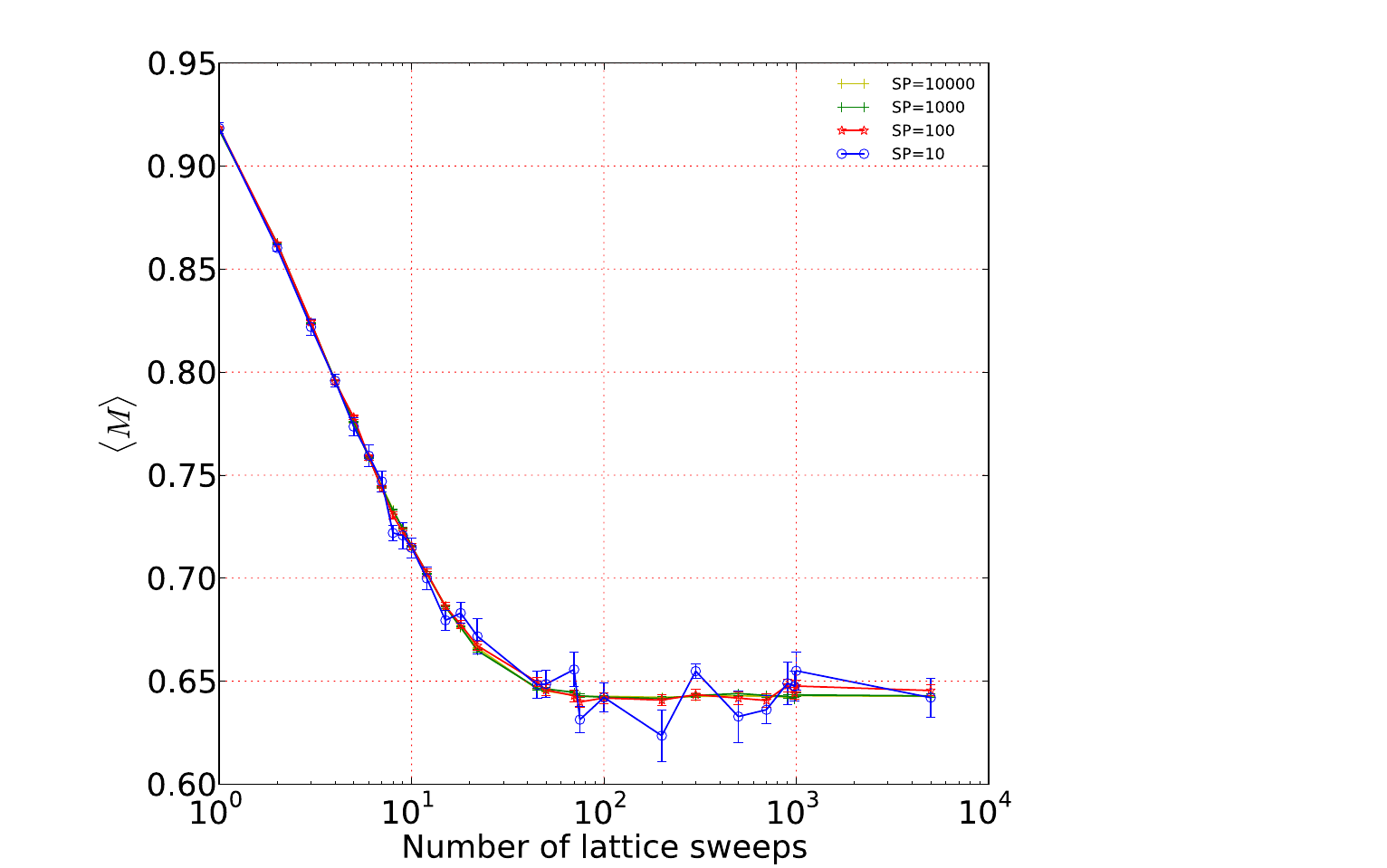}
\centering
\caption{(Color online) Magnetization, calculated starting
from ordered state, as a function
of number of lattice sweeps for different number of simulation paths, at $T=1$, for
$\mathcal{N}=10,100,1000$ and $10000$.
}
\label{Fig:eee}
\end{figure}

The number of lattice sweeps needed for a lattice to reach it's representative state 
(also called burn-in or warm--up phase) is unknown.
It depends on many parameters and can vary substantially.
Insufficient number of lattice sweeps causes inaccurate simulation results.
To overcome this problem for each temperature
half of the simulation paths are computed from the random initial state 
where other half started from the ordered state 
These two sets are averaged using (\ref{MC}) but results from each half separately.
When both halves produce the same result (Figure \ref{Fig:ee}) we can be reasonably certain
that it is an accurate value.

\begin{figure}[H] 
\includegraphics [width=14.0cm]{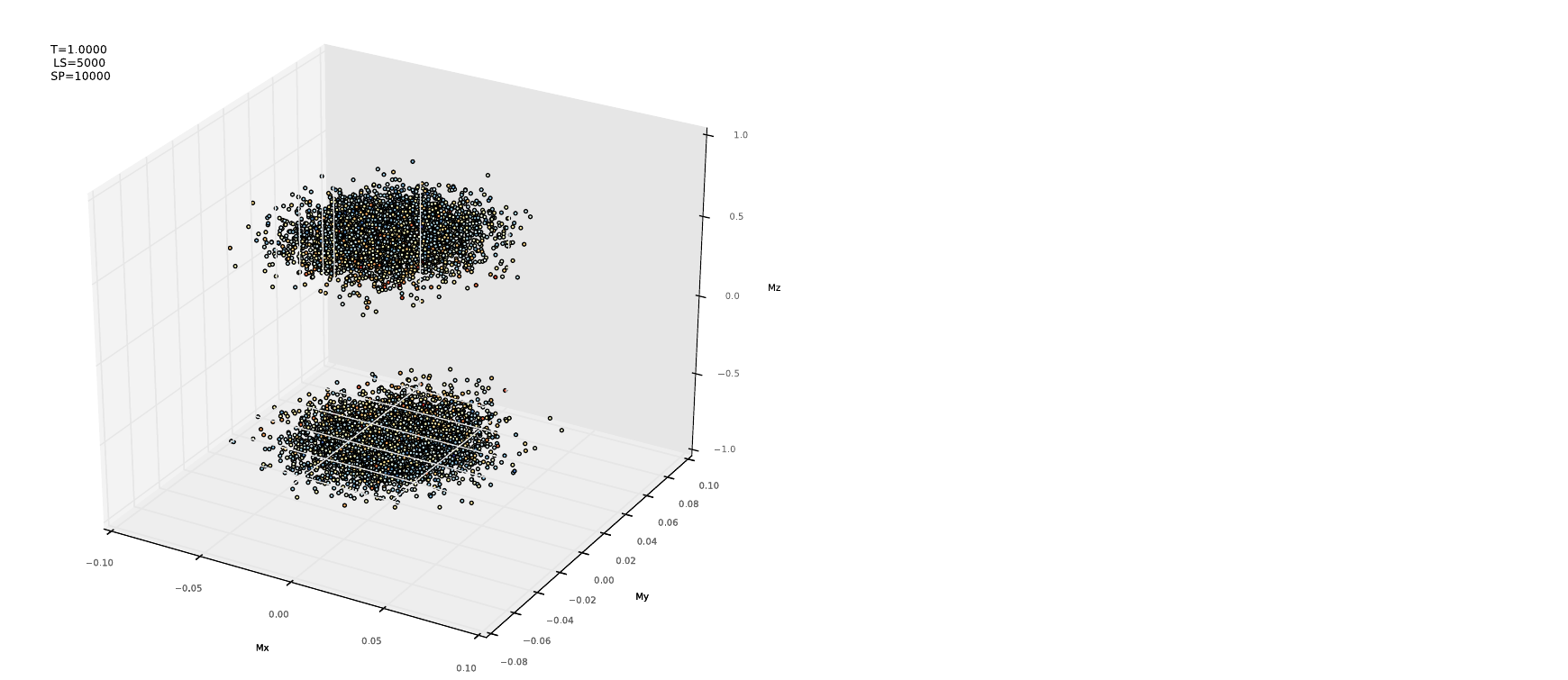}
\centering
\caption{(Color online) Distribution of total spin 
at $T=1$, for $5\times 10^3$ lattice sweeps and $10^4$ simulation paths. Every path
started from random spin configuration, where both angles
$\theta$ and $\varphi$ are taken to be random. }
\label{Fig:goredole}
\end{figure}

Total spin distribution at $T=1$, with
$5\times 10^3$ lattice sweeps and $10000$ simulation paths is presented in Figure \ref{Fig:goredole}. In
Figure \ref{Fig:goredole} every path starts from  random lattice configuration. From all those measurments magnetization
is obtained (see gray line at Figure \ref{Fig:ee}).

\begin{figure}[hbt] 
\includegraphics [width=7.0cm]{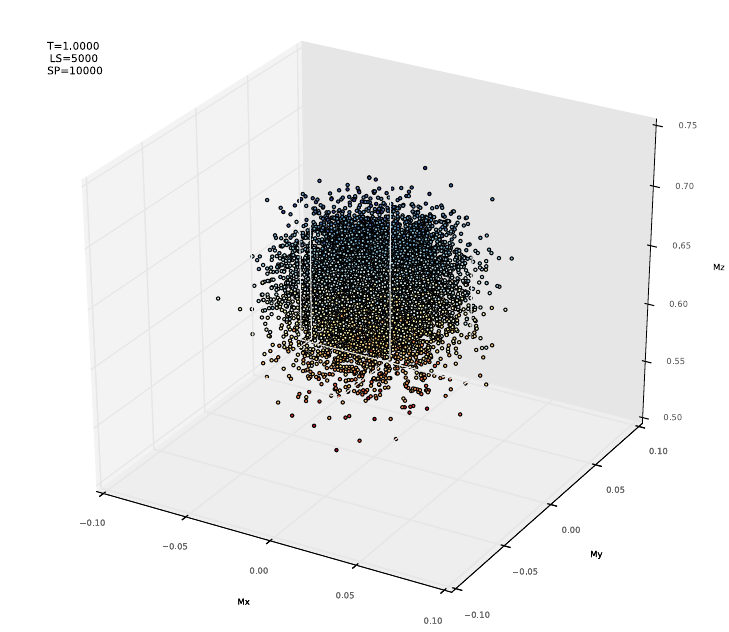}
\centering
\caption{(Color online)  Distribution of total spin 
at $T=1$, for $5\times 10^3$ lattice sweeps and $10^4$ simulation paths. 
Every path started from ordered configuration, with $\theta=0$ and
$\varphi$ arbitrary. }
\label{Fig:gore}
\end{figure}

\begin{figure}[hbt] 
\includegraphics [width=13.0cm]{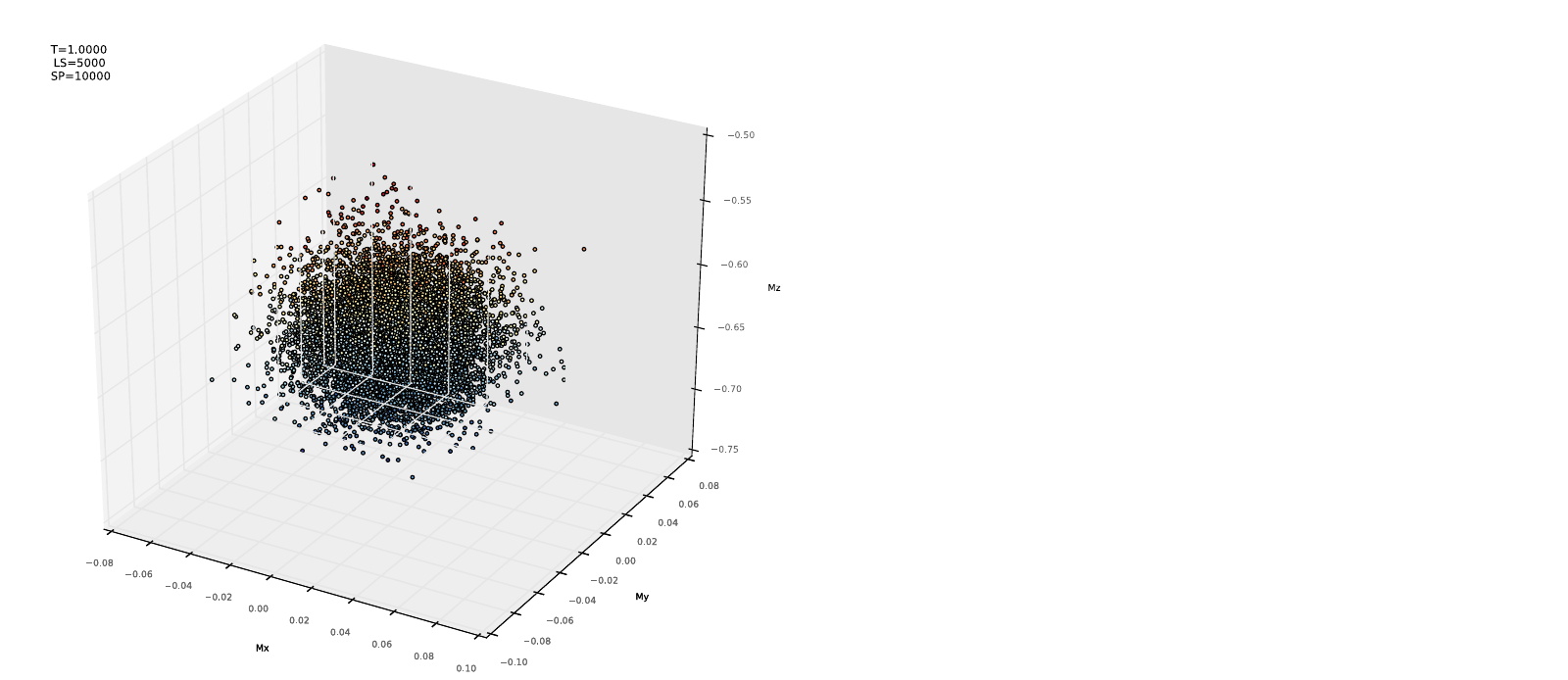}
\centering
\caption{(Color online) Distribution of total spin at 
$T=1$, for $5\times 10^3$ lattice sweeps and 10000 simulation paths.
Every path started from ordered configuration, with $\theta=\pi$ and
$\varphi$ arbitrary. }
\label{Fig:dole}
\end{figure}

However, contrary of that, total spin
distributions in Figs \ref{Fig:gore} and \ref{Fig:dole} are
obtained from multipath simulation where every path started
from ordered state. Both sets of measurments,
 one from Figure \ref{Fig:gore}
and the other one from Figure \ref{Fig:dole}, give the same value of magnetization
(blue line in Figure \ref{Fig:ee}).

 Multipath approach of the $O(3)$ classical Heisenberg model
shows phase transition from the ordered ferromagnetic phase to the
 paramagnetic phase at temperature $T_c=1.442(20)$ (see \cite{nas}). 
%

To demonstrate the applicability of multipath approach we examined the 
thermodynamical properties of classical Heisenberg model and 
compared it with the results obtained from conventional single--path approach.
As expected,the results are in good agreement.
The multipath approach produces statistically independent results 
on which standard statistical methods can be applied \cite{nas}.
Therefore, it is possible to conduct a ''short simulation''
for a quick qualitative analysis (Figure \ref{Fig:eee}), 
which can be of great importance in research of new models.

\section*{Acknowledgments}
This work was supported by the Serbian Ministry of
Education and Science under Contract No. OI-171009.
The authors acknowledge the use of the Computer Cluster of the
Galicia Supercomputing Centre (CESGA).

\section*{References}
\bibliographystyle{elsarticle-num}
\bibliography{paper2}

\end{document}